\documentclass[12pt]{article}

\usepackage{amsmath,amsfonts,amssymb}
\usepackage{psfrag}
\usepackage{enumerate}
\usepackage{mathrsfs}
\usepackage{graphicx}
\usepackage{wrapfig}
\usepackage{xcolor}
\usepackage{caption}
\newcommand\blfootnote[1]{%
  \begingroup
  \renewcommand\thefootnote{}\footnote{\hspace{-6mm}#1}%
  \addtocounter{footnote}{-1}%
  \endgroup
}

\makeatletter\renewcommand\section{\@startsection {section}{1}{\z@}%
                                   {-3.5ex \@plus -1ex \@minus -.2ex}
                                   {2.3ex \@plus.2ex}%
                                   {\normalfont\large\bfseries}}
\renewcommand\subsection{\@startsection{subsection}{2}{\z@}%
                                     {-3.25ex\@plus -1ex \@minus -.2ex}%
                                     {1.5ex \@plus .2ex}%
                                     {\normalfont\bfseries}}

\newcommand{\be}{\begin{equation}}
\newcommand{\ee}{\end{equation}}
\newcommand{\beq}{\begin{eqnarray}}
\newcommand{\eeq}{\end{eqnarray}}

\def\[{\left [}
\def\]{\right ]}
\def\({\left (}
\def\){\right )}

\def\r2{\sqrt{2}}

\newcommand{\bbibitem}[1]{\bibitem{#1}\marginpar{#1}}

\def\Label#1{\label{#1}%
  \smash{\hbox to0pt{\raise1ex\hbox{\tiny[#1]}\hss}}}
\def\noLabels{\let\Label=\label}
\def\nobbibitem{\let\bbibitem=\bibitem}

\begin{document}
\noLabels 
\nobbibitem 

\clearpage\thispagestyle{empty}
\begin{center}
{\Large \bf  Entwinement and the emergence of spacetime}

\vspace{7mm}

Vijay Balasubramanian$^{a,b}$, Borun D. Chowdhury$^{c}$, Bart{\l}omiej Czech$^{d}$, \\
Jan de Boer$^{e}$

\blfootnote{\tt vijay@physics.upenn.edu,czech@stanford.edu,bdchowdh@asu.edu,J.deBoer@uva.nl}


\bigskip\centerline{$^a$\it David Rittenhouse Laboratories, University of Pennsylvania}
\smallskip\centerline{\it 209 S 33$^{\rm rd}$ Street, Philadelphia, PA 19104, USA}
\bigskip\centerline{$^b$\it CUNY Graduate Center, Initiative for the Theoretical Sciences}
\smallskip\centerline{\it 365 Fifth Avenue, New York, NY 10016, USA}
\bigskip\centerline{$^c$\it Department of Physics, Arizona State University}
\smallskip\centerline{\it Tempe, Arizona 85287, USA}
\bigskip\centerline{$^d$\it Department of Physics, Stanford University}
\smallskip\centerline{\it 382 Via Pueblo Mall, Stanford, CA 94305-4060, USA}
\bigskip\centerline{$^e$\it Institute for Theoretical Physics, University of Amsterdam}
\smallskip\centerline{\it Science Park 904, Postbus 94485, 1090 GL Amsterdam, The Netherlands}
\end{center}

\vspace{5mm}

\begin{abstract}
\noindent
It is conventional to study the entanglement between spatial regions of a quantum field theory.   However, in some systems entanglement can be dominated by ``internal'', possibly gauged, degrees of freedom that are not spatially organized, and that can give rise to gaps smaller than the inverse size of the system.   In a holographic context, such small gaps are associated to the appearance of horizons and singularities in the dual spacetime.   Here, we propose a concept of {\it entwinement}, which is intended to capture this fine structure of the wavefunction. Holographically, entwinement probes the {\it entanglement shadow} -- the region of spacetime not probed by the minimal surfaces that compute spatial entanglement in the dual field theory.    We consider the simplest example of this scenario -- a 2d  conformal field theory (CFT) that is dual to a conical defect in AdS$_3$ space.     Following our previous work, we show that spatial entanglement in the CFT reproduces spacetime geometry up to a finite distance from the conical defect.     We then show that the interior geometry up to the defect can be reconstructed from entwinement that is sensitive to the discretely gauged, fractionated degrees of freedom of the CFT.        Entwinement in the  CFT is related to non-minimal geodesics in the conical defect geometry, suggesting a potential quantum information theoretic meaning for these objects in a holographic context.   These results may be relevant for the reconstruction of black hole interiors from a dual field theory.
\end{abstract}

\setcounter{footnote}{0}
\newpage
\clearpage
\setcounter{page}{1}

\section{Introduction}

According to the Ryu-Takayanagi (RT) proposal \cite{rt, RT},  classical geometry and quantum entanglement are  related via holographic duality. The proposal states that the entanglement entropy of a spatial region $\mathcal{R}$ in the field theory is given by:
\begin{equation}
S(\mathcal{R}) = \frac{1}{4G} 
\min_{\partial \mathcal{A} = \partial\mathcal{R}} {\rm Area}(\mathcal{A}). \label{eqrt}
\end{equation}
In this formula, which assumes that the bulk spacetime is static,\footnote{See \cite{HRT} for a generalization to non-static spacetimes.} the minimum is taken over 
bulk surfaces, which are contained in the same spatial slice as the boundary region $\mathcal{R}$ and which asymptote to the boundary of $\mathcal{R}$. 
The RT formula relates entanglement entropy, a non-local quantity in the boundary theory, to a minimal surface, which is a nonlocal object in the bulk.
  Recently, we proposed a new quantity, the {\it differential entropy}, constructed out of entanglement, that reconstructs the areas of closed surfaces in AdS that do not asymptote to the boundary \cite{holeinspacetime,lastpaper,myersetal, xijamieandi}.\footnote{This quantity was found while trying to make the proposal of \cite{Bianchi:2012ev}, that areas of general surfaces in spacetime are directly related to entanglement across them, more precise.}   By shrinking such closed surfaces one can attempt to reconstruct local geometry in AdS space from purely field theoretic objects \cite{recon1, recon2, recon3, recon4, lamprosczech}. The relevance of boundary entanglement for such a reconstruction was first pointed out in \cite{marksessay, brian1, brian2, Bianchi:2012ev}, see also \cite{erepr}.

In order for this program to succeed, it is necessary for the union of RT minimal surfaces to cover all of spacetime.  This is barely possible in empty AdS space where the largest RT surfaces, associated to the entanglement of half of the field theory, are necessary to include the origin of space.   But away from pure AdS, an {\it entanglement shadow} can develop -- a region which is not probed by minimal surfaces and hence by conventional spatial entanglement entropy in the dual field theory.
(Our terminology is inspired by the term {\it causal shadow} introduced by \cite{mattetal} to describe a region which is causally disconnected from all the spacetime boundaries.)   For example, the AdS-Schwarzschild and BTZ black holes have an entanglement shadow of thickness of order the AdS scale surrounding the horizon \cite{plateaux}.
 This conundrum reflects a general difficulty in the AdS/CFT correspondence of identifying field theoretic observables associated to physics in a region of size less than one AdS volume \cite{banks, peetjoe, Balasubramanian:1999ri, giddings, kll, jamiejoe, jaredliam, pr}.  

How can we see inside entanglement shadows?     Many lines of evidence point to the idea that recovering the local physics in such regions involves ``internal'' degrees of freedom of the CFT that are not themselves spatially organized.   Consider, for example, the role of the matrix degrees of freedom in the $SU(N)$ Yang-Mills theory dual to AdS$_5$ \cite{susskindwitten} and the fractionated degrees of freedom in the D1-D5 string dual to AdS$_3$ \cite{juanlenny} in reconstructing the deep interior of space \cite{boussomints}, the compact dimensions of the bulk \cite{llm,berenstein,babel}, and the entropy of AdS black holes \cite{wittenonblackholes, lenny93, stromingervafa, dasmathur, klebanovgubserpeet}.   Such internal degrees of freedom can have energy gaps much smaller than those dictated by the spatial size of the system, and thus represent a deep IR regime of the field theory that will on general grounds be associated to the deep interior of AdS.\footnote{See \cite{vijaymark} for a discussion of entanglement between high and low momenta in a field theory.  We are here discussing entanglement between IR degrees of freedom that are not spatially organized, so we need a different formalism.}

All of this suggests that to see inside entanglement shadows we will need to consider the entanglement of  internal degrees of freedom with each other.  If the Hilbert space for these variables factorizes, we can derive a reduced density matrix for any subset of them and compute its entanglement entropy.  However, there is often an additional subtlety -- typical realizations of holography involve gauge symmetries acting on the internal degrees of freedom.       In this context, how do we ask  questions like ``How entangled is a subset of the degrees of freedom with the rest of the theory?"   The challenge here is that the subset in question may not be gauge invariant by itself.      We propose to deal with this problem in a pedestrian fashion: embed the theory in an auxiliary, larger theory, where the degrees of freedom are not gauged, compute conventional entanglement there, then sum over gauge copies to get a gauge invariant result.   (We will deal with discrete gauge groups in this paper where the sum over gauge copies is easy to define; for continuous gauge symmetries  consideration of an appropriate measure would be necessary.)  This is not a conventional notion of entanglement that is associated to a gauge invariant algebra of observables.  Hence we give it a new name  --  {\it entwinement}.  The justification for inventing this concept is that it will turn out to have a useful meaning in the dual gravity theory in terms of extremal but non-minimal surfaces in the examples we consider.\footnote{A notion of un-gauging or expanding the Hilbert space and then re-gauging has also appeared in \cite{verlinde1, verlinde2}, which discuss the problem of seeing behind a horizon in AdS/CFT.}

A trivial example of entwinement is to consider two CFTs living on the same space. One can compute the entanglement entropy of the degrees of freedom living in a spatial region $\mathcal{R}$ in CFT$_1$ and CFT$_2$ separately, and then sum up the results. This is a rudimentary example of entwinement. When the two CFTs are in a product state, this computation returns the entanglement entropy of the region $\mathcal{R}$, but in a general state entwinement is distinct from entanglement. A different example is to consider matrix degrees of freedom in a local theory, for example in the ${\cal N} = 4$ super-Yang-Mills that is dual to AdS$_5$.    One can imagine computing the entanglement of a subset of the matrix degrees of freedom in some spatial region.  However, because of the gauge symmetry we cannot do this na{\"i}vely. One option is to ungauge and then sum over gauge copies; this would give the entwinement that we propose to define.  In this paper we concentrate on a different example: a $1+1$-dimensional  conformal field theory dual to a conical defect in AdS$_3$ spacetime.  We will see that in this setting, the entwinement of certain fractionated degrees of freedom can be understood in geometric terms by going to the covering space and applying the Ryu-Takayanagi proposal. The upshot is that the entwinement of the fractionated degrees of freedom corresponds in the dual conical defect spacetime to non-minimal geodesics -- curves, which are local but not global minima of the distance function. This analysis comprises Sec.~2.

In Sec.~3, we apply these results to the formalism we developed in \cite{lastpaper}, which reconstructs analytically the bulk geometry from field theory data.   We find that the conical defect spacetime contains a macroscopically large entanglement shadow -- a central zone surrounding the conical defect that cannot be reconstructed from entanglement entropies of spatial boundary regions.  The geometry in the central zone is determined by entwinement, as we explicitly demonstrate.  In fact, we will show that even outside the entanglement shadow there are geometric objects whose boundary description involves entwinement.    The conical defect spacetime is similar to a black hole in having a macroscopic entanglement shadow whose size is controlled by the mass of the object.    In Sec.~4 we discuss how our findings inform the debate about reconstructing the geometry near and beyond the horizon of a black hole.   

\section{Conical defects, long geodesics and entwinement}

\subsection{The entanglement shadow of the conical defect geometries}
\label{sec:entshadow}
\begin{figure}[t!]
\centering
\includegraphics[width=.41\textwidth]{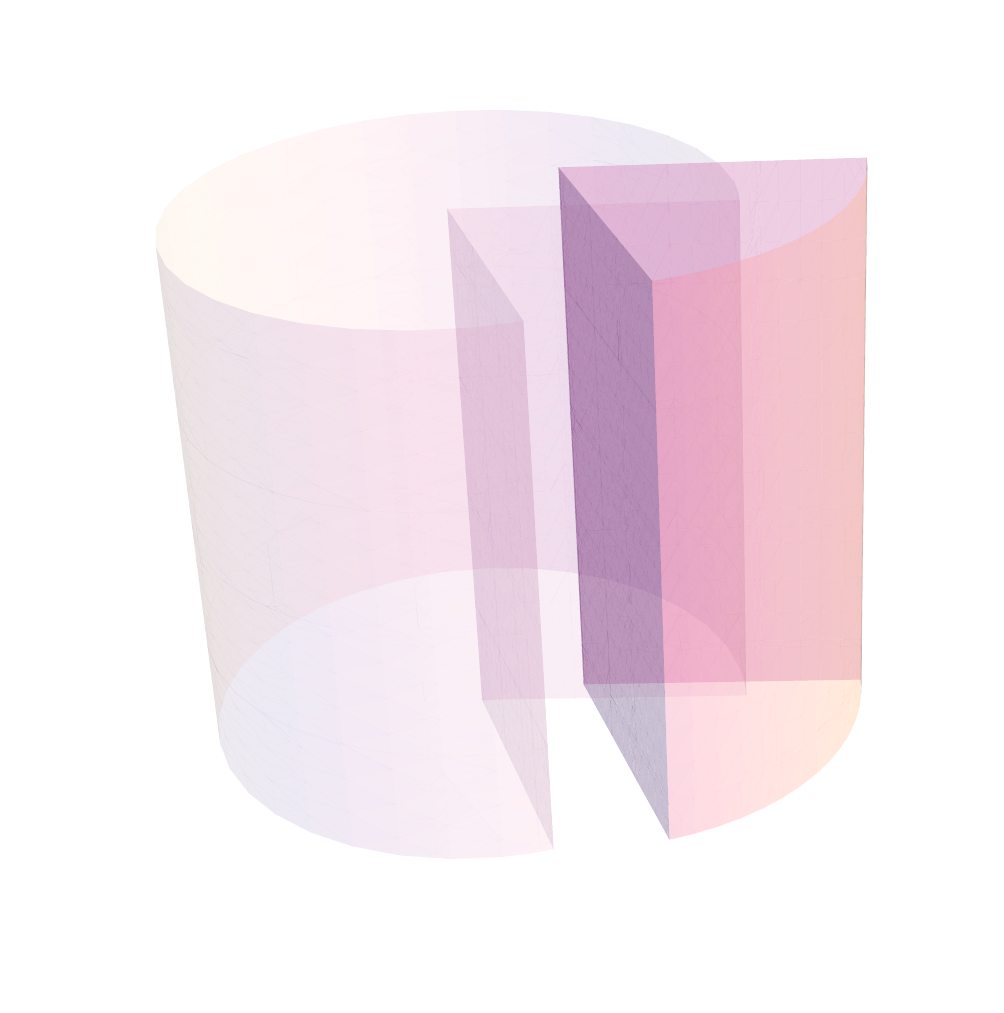}
\caption{A conical defect geometry as a wedge cut out of AdS space.}
\label{adswedge}
\end{figure}

We begin with a review of the conical defect geometry. Starting from the global AdS$_3$ coordinates, 
\begin{equation}
ds^2 = - \left( 1 + \frac{R^2}{L^2}\right) dT^2 + \left( 1 + \frac{R^2}{L^2}\right)^{-1} dR^2 + R^2 d\tilde\theta^2,
\label{covermetric}
\end{equation}
we obtain AdS$_3/\mathbb{Z}_n$ simply by declaring the angular coordinate to be periodic with period $2\pi/n$ as shown in Fig.~\ref{adswedge}. 
Another useful coordinate system is obtained from the first by the rescaling
\begin{equation}
\theta = n \tilde\theta \qquad {\rm and} \qquad r = R / n \qquad {\rm and} \qquad t = n T,
\label{rescalings}
\end{equation}
which leads to:
\begin{equation}
ds^2 = - \left( \frac{1}{n^2} + \frac{r^2}{L^2}\right) dt^2 + \left( \frac{1}{n^2} + \frac{r^2}{L^2}\right)^{-1} dr^2 + r^2 d\theta^2 \,.
\label{defmetric}
\end{equation}
In these coordinates, $\theta$ ranges between 0 and $2 \pi$. The deficit angle around $R = r = 0$ means that the spacetime is singular on this locus.

\begin{figure}[t!]
\centering
\begin{tabular}{ccc}
\includegraphics[width=.34\textwidth]{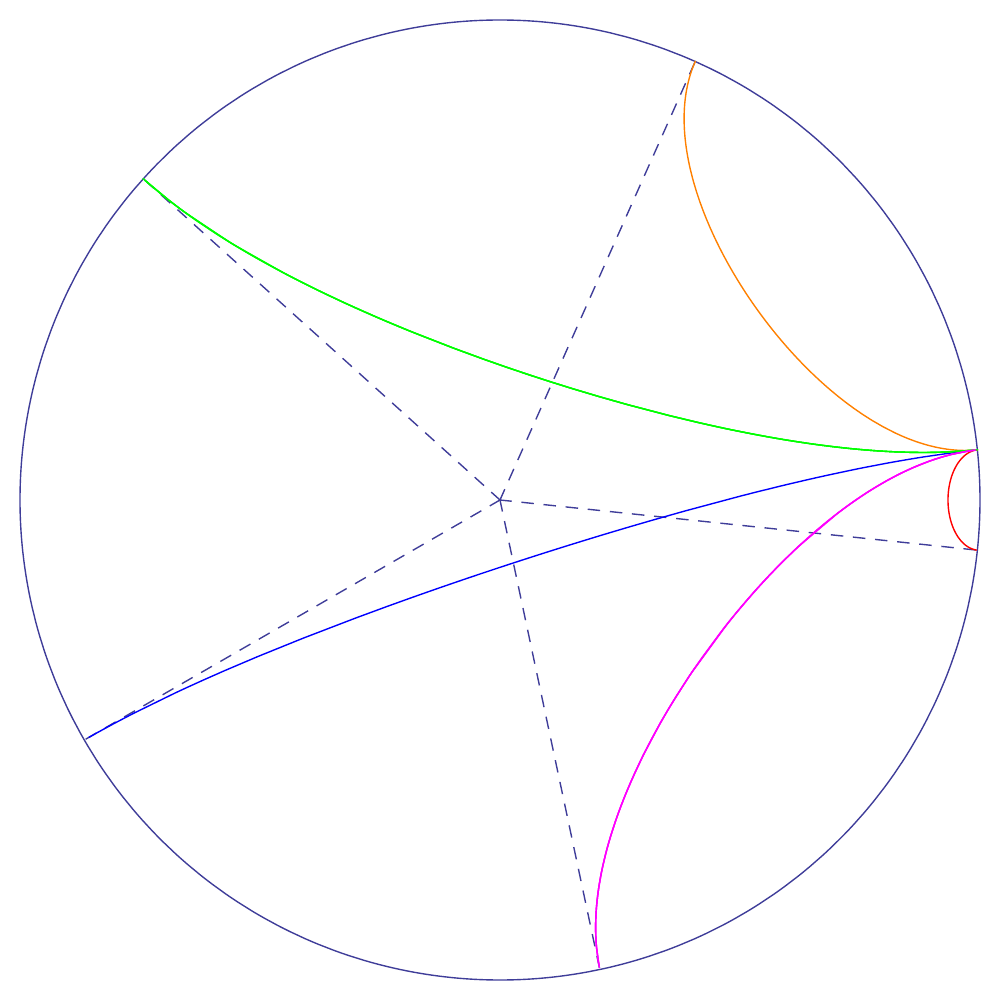} & \qquad\qquad\qquad &
\includegraphics[width=.34\textwidth]{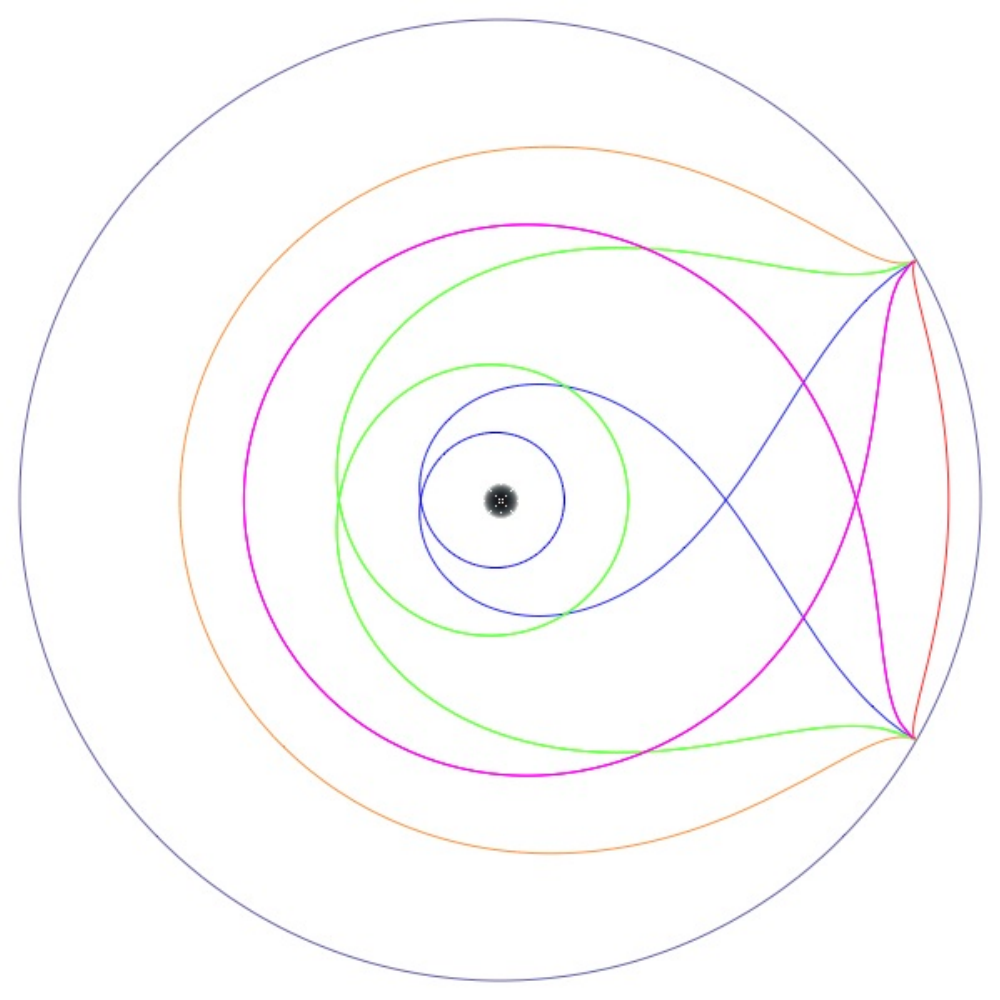}
\end{tabular}
\caption{A spatial slice of anti-de Sitter space (left) and of the conical defect ${\rm AdS }_3/ \mathbb{Z}_n$ (right). Spatial geodesics in the conical defect geometry descend from geodesics in the covering space with one endpoint ranging over the $n$ images. All but one of them are long geodesics.}
\label{geodesicspic}
\end{figure}

\paragraph{Spatial geodesics}
Using the coordinate transformation~(\ref{rescalings}), it is trivial to find the spatial geodesics in the conical defect spacetime. One starts with the $n$ geodesics in the AdS geometry~(\ref{covermetric}), which have one endpoint in common while the other ranges over a family of $n$ points that are $2\pi / n$ apart from one another. After the identification by $\tilde\theta \cong \tilde\theta + 2\pi / n$, these geodesics descend to a family of $n$ distinct geodesics with the same endpoints in the conical defect geometry. This is illustrated in Fig.~\ref{geodesicspic}. The geodesics are described by the equations
\begin{equation}
\tan^2\tilde\theta = \frac{R^2 \tan^2\tilde\alpha - L^2}{R^2 + L^2} 
\qquad \Leftrightarrow \qquad
\tan^2(\theta/n) = \frac{n^2 r^2 \tan^2(\alpha/n) - L^2}{n^2 r^2 + L^2} 
\label{defgeodesics}
\end{equation}
and their length is:
\begin{equation}
l(\alpha) = 
2L \log \left( \frac{2L}{\mu} \sin\tilde\alpha \right) = 
2L \log \left( \frac{2L}{\mu} \sin(\alpha/n) \right).
\label{deflength}
\end{equation}
In our notation the opening angle in the coordinates~(\ref{defmetric}) is $\alpha$ while the opening angle in the covering coordinates~(\ref{covermetric}) is $\tilde\alpha = \alpha / n$.   By opening angle, we mean half the angular size of the boundary interval subsumed by the geodesic, including winding.   For example, a geodesic that winds twice completely around the conical defect and returns to the original point would subsume a boundary angle $4\pi$ and have $\alpha = 2\pi$.  In  this language, the longest geodesic has $\alpha = n\pi / 2$.   The quantity $\mu^{-1}$ is a gravitational IR regulator, which cuts off the infinite tails of the geodesics near the spacetime boundary.\footnote{If we can compare cutoffs by
matching radial positions in a standard Fefferman-Graham expansion near infinity, then perhaps
one should rescale the cutoff with a factor of $n$ as well.   Such a rescaling  would yield a simple additive contribution to the entanglement entropy which we will ignore in the remainder.}

The minimal geodesics~(\ref{defgeodesics})  between any pair of points have  $\alpha \leq \pi/2$; thus they compute entanglement entropies of spatial intervals of angular size $2\alpha \leq \pi$ in the dual CFT \cite{rt, RT}.  The spatial entanglement of intervals of angular size $\pi \leq 2\alpha \leq 2\pi$ is equal to that of the complementary interval of angular size $2\pi - 2\alpha$.   This is automatic if the geometry describes a pure state.   Even if there is a mixed state with the same geometric description, it would not have macroscopic entropy and  the associated ``horizon'' would have to hug the conical defect, and hence the RT prescription, which in this case includes the horizon area, would still give the same answer.

The minimal geodesics (with $\alpha \leq \pi/2$) penetrate the bulk up to the radial location
\begin{equation}
r_{\rm crit}(n) = \frac{L \cot(\pi/2n)}{n}\,. 
\label{defmin}
\end{equation}
Thus, the central zone $r < r_{\rm crit}(n)$ is not probed at all by entanglement entropies of spatial regions of the boundary. We call this zone of spacetime the \emph{entanglement shadow}.

The  $n-1$ {\it long geodesics} between a given pair of points on the boundary do not have the interpretation of computing entanglement entropy, because they do not satisfy the minimality condition in formula~(\ref{eqrt}).   All of these geodesics have $\alpha > \pi/2$, and will turn out to compute entwinement -- the novel concept that is the subject of the present paper.   In Sec.~3 we will show how the full conical defect geometry, including the entanglement shadow, can be reconstructed in the formalism of \cite{lastpaper} using boundary entwinement as input.

\subsection{Ungauging the dual description of the conical defect}
\label{orbifoldcft}

There are various ways of representing the conical defect spacetime in a dual field theory.  One approach is to regard the defect as an excited state of AdS$_3$.   In this picture, we start with a conformal field theory CFT$_c$ of central charge $c = 3L/2G$ where $L$ is the AdS scale and $G$ is the Newton constant in three dimensions.  The vacuum of this theory is empty AdS space and the conical defect is a particular excited state.   We will discuss this description further below, but  turn at present to a more convenient view of the system in terms of its covering space.

  As described above, AdS$_3/\mathbb{Z}_n$ can be regarded as an angular identification of a covering AdS$_3$ spacetime.   The covering space ``ungauges'' the $\mathbb{Z}_n$ discrete gauge symmetry and physical quantities are computed by considering $\mathbb{Z}_n$-invariant quantitites in the ungauged theory.    Indeed, the correlation functions of quantum fields in the conical defect and BTZ spacetimes are typically computed precisely by taking this sort of view, which is equivalent to the method of images for computing Green functions \cite{eskoBTZ,vijaymasaki}.   Boundary limits of these Green functions correctly compute the correlation functions of the corresponding CFT states \cite{simonvijay, vijaymasaki}.    
This does not say  that AdS$_3$ with a conical defect is exactly identical to the covering theory in its ground state;
in fact it is not. However, many $\mathbb{Z}_n$ invariant observables and correlation functions computed in the covering theory agree with their corresponding counterparts in the conical defect theory.

   From this perspective, the field theory dual to the conical defect should also be lifted to the covering space, which is an $n$-times longer circle.  We will denote  this parent theory CFT$_{\tilde{c}}$, where $\tilde{c}$ is a new central charge to be determined later.   Spatial locations $x$ in CFT$_c$ lift to locations $\tilde{x}$ in a fundamental domain of the covering space and to the corresponding $\mathbb{Z}_n$ translates.    Correlations functions of  CFT$_{\tilde{c}}$  that descend to CFT$_c$ must be $\mathbb{Z}_n$-symmetric and are computed by symmetrized operators $\mathcal{O} = \sum_{i=0}^{n-1} g^i \tilde{\mathcal{O}}$, where $\tilde{\mathcal{O}}$ is any CFT$_{\tilde{c}}$ operator and $g$ is a $\mathbb{Z}_n$ generator.   This recapitulates the method of images used in the bulk to compute the same correlation functions \cite{vijaymasaki}.  In the geodesic approximation \cite{simonvijay} the correlation functions between $\mathcal{O}_1(x)$ and $\mathcal{O}_2(y)$ would be computed from the geodesics between lifts $(\tilde{x},\tilde{y})$ of $(x,y)$ to the covering space, and between all $\mathbb{Z}_n$ translations of these locations.    The geodesic between $\tilde{x}$ and $\tilde{y}$ in the covering space descends to the minimal geodesic on the defect, and the geodesics between $\tilde{x}$ and the $\mathbb{Z}_n$ translates of $\tilde{y}$ descend to the long geodesics on the defect geometry.   The leading contribution to the correlator comes from the minimal geodesics in Fig.~\ref{geodesicspic}, but the long geodesics yield subleading saddle points  and are all necessary to give the correct correlation function in the defect theory.

What is the central charge $\tilde{c}$ of the covering CFT?   We will give three arguments that $\tilde{c} = c/n$.  First, recall the Brown-Henneaux construction of the asymptotic symmetry algebra of the AdS$_3$ spacetimes \cite{BrownHenneaux}.  In this construction the central charge $\tilde{c}$ of the covering space is derived from the Virasoro algebra of large diffeomorphisms of spacetime:
\begin{equation}
[\tilde{L}_k,\tilde{L}_s] = (s-k) \tilde{L}_{k+s} + {\tilde{c} \over 12} k(k^2 - 1) \delta_{k+s} \,. 
\end{equation}
 But not all such diffeomorphisms of the covering space will descend to the defect theory.  To preserve $\mathbb{Z}_n$ symmetry we must restrict to a subalgebra generated by $\tilde{L}_{nk}$, because $\tilde{L}_k$ are Fourier modes of the boundary deformations.   Following \cite{JanJoanShahin} we recognize that this subalgebra also has Virasoro form with the following redefinition of the generators:
 \begin{equation}
 L_k = {1\over n} \tilde{L}_{nk}, \ \ k \neq 0 ~~~~~~~~~~~~~ L_0 = {1 \over n} \left(\tilde{L}_0 - {\tilde{c} \over 24} \right) + {n\, \tilde{c} \over 24}  
 \end{equation}
This is the Virasoro algebra of deformations that descend to the defect theory; it has a central charge $c= n \, \tilde{c}$.   Since the central charge is related to bulk parameters as $c = 3 L / 2G$, we will interpret this relation as also saying that the covering theory has a rescaled gravitational coupling $\tilde{G} = n G$.  The spectrum of $L_0$ is rescaled by $1/n$ relative to $\tilde{L}_0$.  This indicates that the CFT$_c$ dual to the defect theory has a fractionated spectrum, where momenta are quantized in units of $1/n$ times the length of the spatial circle.   Below we will see direct evidence for this in a weak coupling limit of the CFT, where it can be visualized in terms of an $n$-wound string.

\begin{figure}[t!]
\centering
\begin{tabular}{ccc}
\includegraphics[width=.25\textwidth]{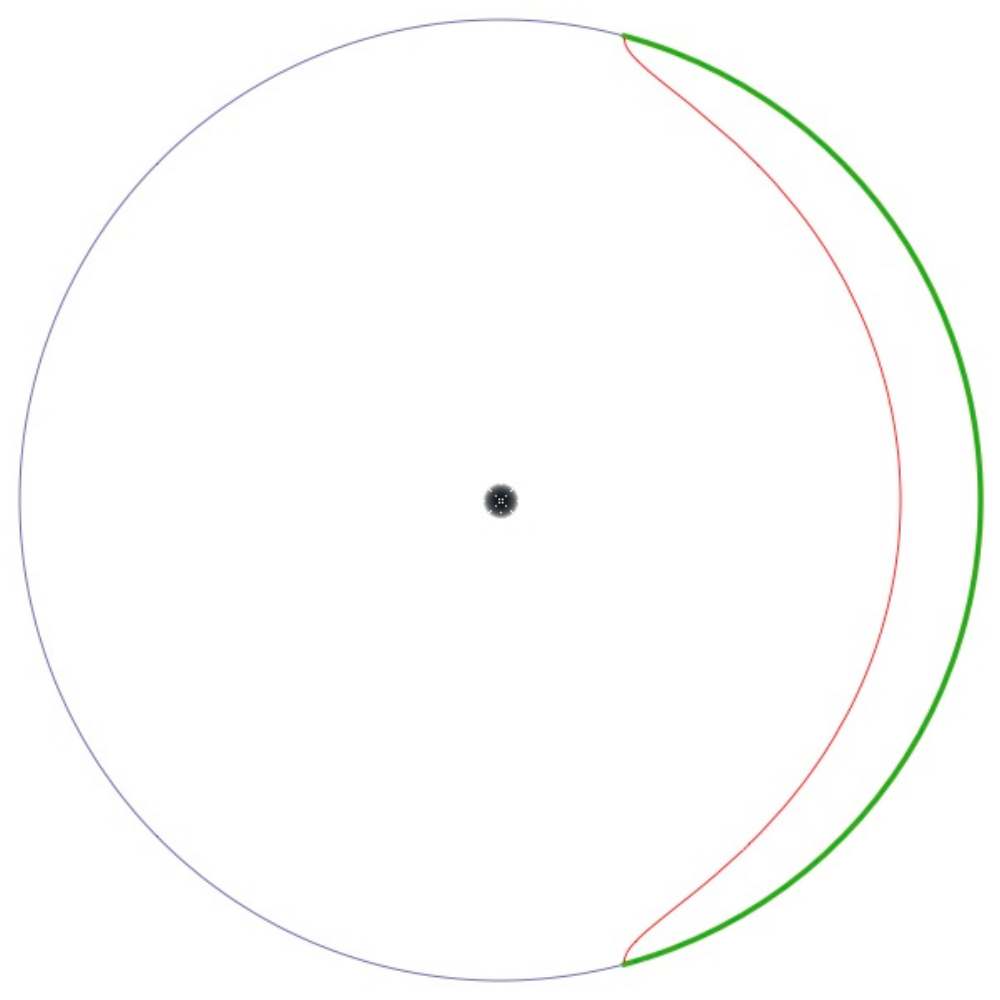} & \qquad\qquad\qquad &
\includegraphics[width=.25\textwidth]{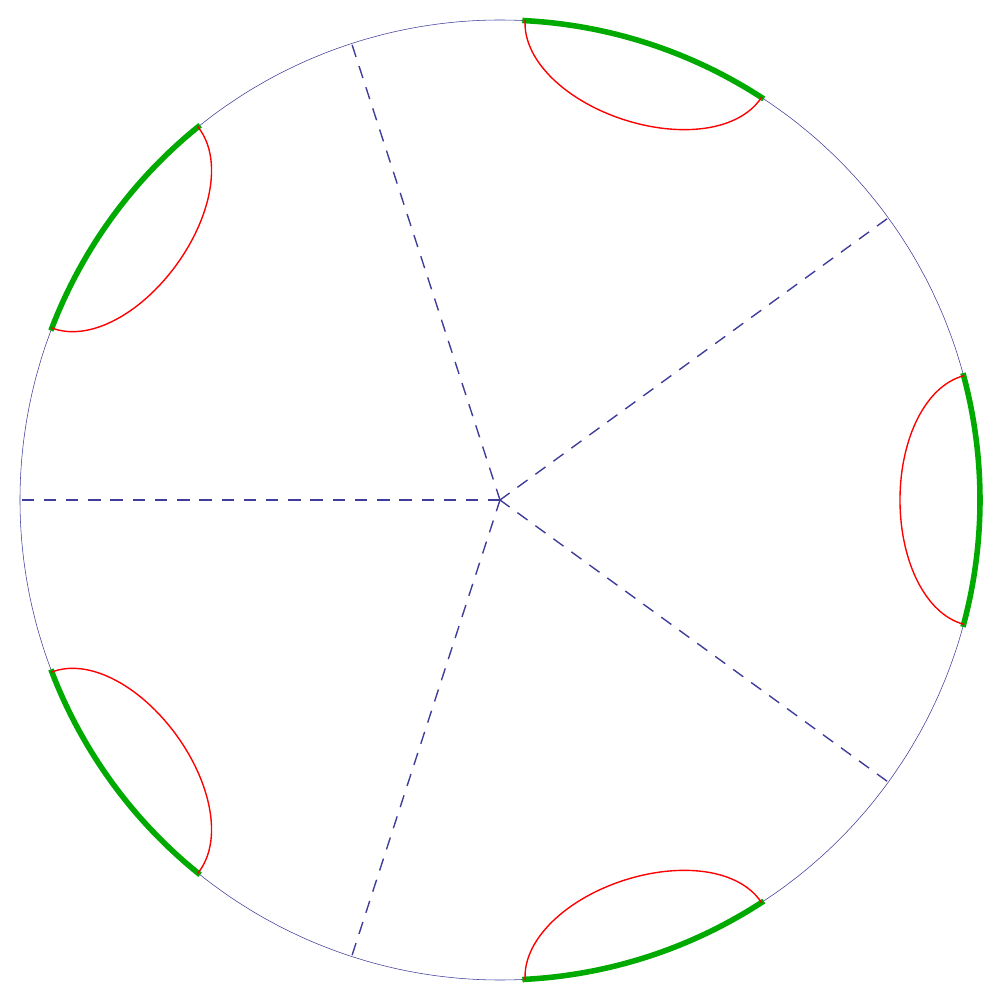} \\
& \\
\includegraphics[width=.25\textwidth]{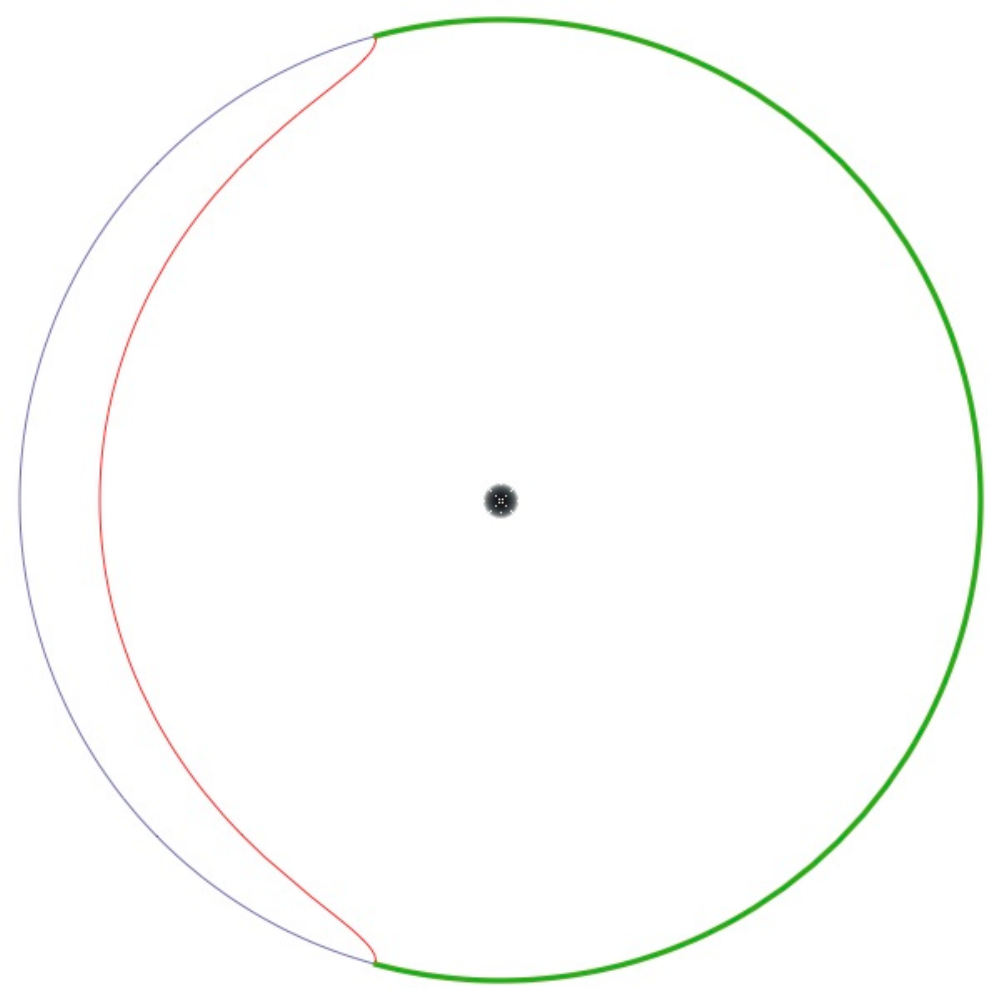} & \qquad\qquad\qquad &
\includegraphics[width=.25\textwidth]{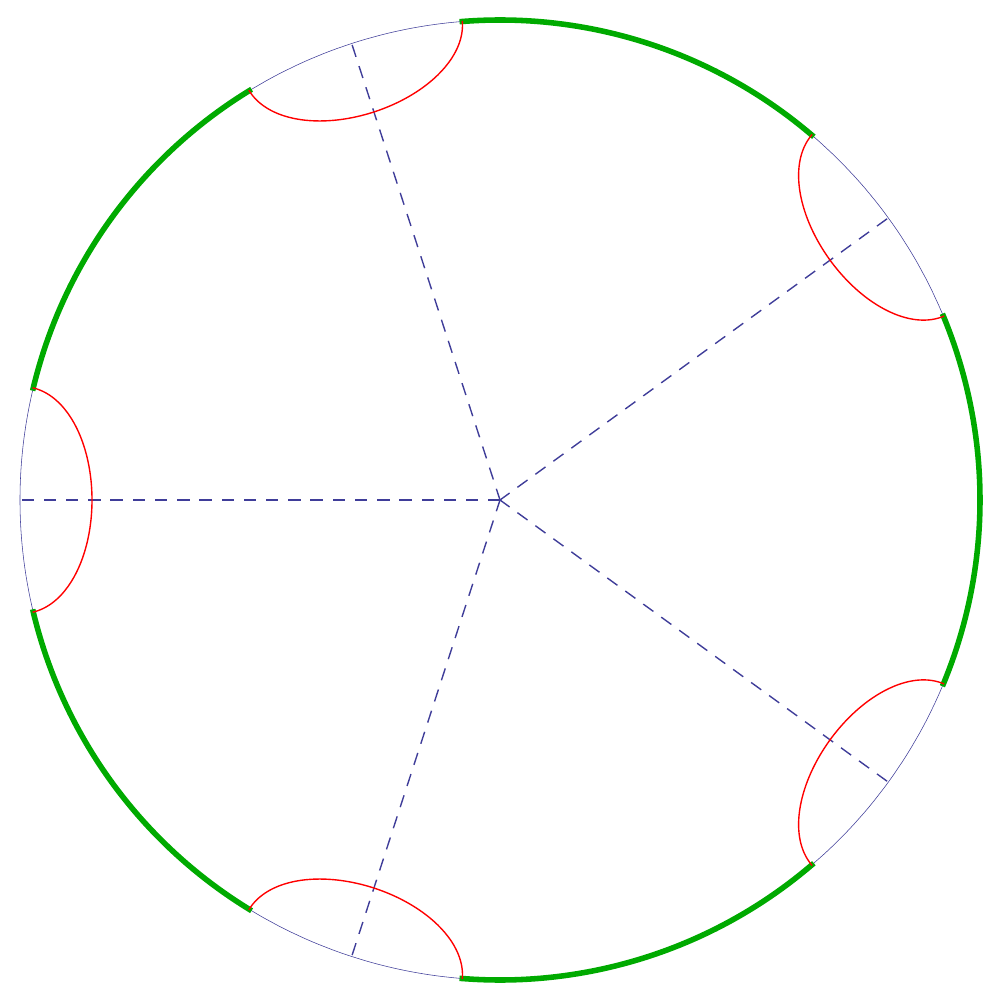} \\
\end{tabular}
\caption{The holographic computation of the entanglement entropy of an interval (green) of width $2\alpha < \pi$ (above) and $2\alpha > \pi$ (below), shown in the short string picture (left) and in the long string picture (right).   The short string interval maps to a union of disjoint long string intervals.  The geometries on the left represent the spatial slice of the conical defect while the geometries on the right are spatial slices of anti-de Sitter space, which is the $n$-fold cover of the conical defect.  There is a transition in the shortest geodesic homologous to the boundary interval when $2\alpha = \pi$ .   Here $n=5$. }
\label{rtincover}
\end{figure}

To test our identification of $\tilde{c}$ let us compare entanglement entropies computed directly in CFT$_c$ and from CFT$_{\tilde{c}}$.   Consider an interval $\mathcal{R}$ in CFT$_c$ with angular size $2\alpha < \pi$.    Applying the Ryu-Takayanagi proposal~(\ref{eqrt}) gives:
\begin{equation}
S(\mathcal{R}) = \frac{1}{4G}\, l(\alpha) = 
\frac{c}{3} \log \sin\frac{\alpha}{n} + {\rm const.}
\label{shortcal}
\end{equation}
We have substituted eq.~(\ref{deflength}) and the standard relation $c = 3L / 2G$.    If the interval $\mathcal{R}$ is larger than half the field theory circle, $2\alpha > \pi$, the entanglement entropy is:
\begin{equation}
S(\mathcal{R}) = \frac{1}{4G} \, l(\pi-\alpha)
= \frac{c}{3} \log \sin\frac{\pi-\alpha}{n} + {\rm const.}
\label{longercal}
\end{equation}
The relevant geodesics in the conical defect spacetime are displayed in the left panel of Fig.~\ref{rtincover}.

The interval $\mathcal{R}$ in CFT$_c$ lifts to $n$ evenly spaced intervals $\tilde{\mathcal{R}}_i$, each of angular size $2\tilde\alpha = 2\alpha/n$.  The Ryu-Takayanagi formula in CFT$_{\tilde{c}}$ now tells us that the entanglement entropy of this union of intervals is computed from the length of the minimal curve in empty AdS$_3$, which is homologous to their union.  
As shown in the right panel of Fig.~\ref{rtincover}, the minimal surface homologous to the union $\cup_{i=1}^n \tilde{\mathcal{R}}_i$ consists of $n$ geodesics, each of which subtends the angle $2\alpha/n$ or $2(\pi-\alpha)/n$, depending on whether $2\alpha \lessgtr \pi$.    The final result for $2\alpha < \pi$ is the sum of $n$ disjoint geodesic lengths and reads
\begin{equation}
S\left( \cup_{i=1}^n \tilde{\mathcal{R}}_i \right)
= n \cdot \frac{\tilde{c}}{3} \log \sin\tilde{\alpha} + {\rm const.} = \frac{c}{3} \log \sin\frac{\alpha}{n} + {\rm const.} = S(\mathcal{R}),
\label{bothcals}
\end{equation}
with $\alpha \to \pi-\alpha$ in the opposite case.  The relevant geodesics appear in families of $n$ identical images, which guarantees $\mathbb{Z}_n$ invariance of the result.   Here, $\mathbb{Z}_n$-invariance is obtained, because the input to the holographic calculation, namely the union of intervals $\cup_{i=1}^n \tilde{\mathcal{R}}_i$, is by construction symmetrized.   From the perspective of the covering theory, the transition between the geodesics that subtend $2\alpha/n$ or $2(\pi-\alpha)/n$ marks two phases where  disjoint intervals $\tilde{\mathcal{R}}_{i}$ and $\tilde{\mathcal{R}}_{i+1}$ do or do not share mutual information \cite{HeadrickTransition}. Interestingly, this mutual information transition in the CFT$_{\tilde{c}}$ conspires to correctly reproduce the entanglement computations for a single interval in the defect theory, where there is no mutual information to account for.  

   The weakly-coupled limit of the CFT dual to the conical defect is also very instructive concerning the above points.  To arrive at these insights, we first give a lightning review of the relevant facts about the duality relating asymptotically AdS$_3$ spacetimes and the D1-D5 field theory. Consider $N_1$ D1-branes wrapped on an $S^1$ and $N_5$ D5-branes wrapped on $S^1 \times T^4$. The low energy description of this system is a 2d CFT, whose moduli space contains the so-called orbifold point, where the dynamics reduces to a free $\mathcal{N} = (4,4)$ supersymmetric sigma model with target space $(T^4)^N / S_N$ with $N = N_1 N_5$. The near-horizon limit of the geometry sourced by this D-brane system is ${\rm AdS}_3 \times S^3 \times T^4$, with the AdS curvature scale $L$ proportional to $N$ in three-dimensional Planck units.   A weakly coupled type IIB string theory on this geometry is dual to a certain marginal deformation of the orbifold CFT with large $N$. The weakly coupled limit of the CFT is near the orbifold point and corresponds to a strongly coupled AdS theory.  We are going to consider this limit.        The low-energy CFT describing the brane dynamics is identified with the theory dual to ${\rm AdS}_3$, living on the conformal boundary of  this space:
\begin{equation}
ds_\partial^2 = -dt^2 + L^2 d\theta^2 \qquad {\rm with} \qquad \theta \sim \theta + 2\pi.
\label{boundarygeom}
\end{equation}
The central charge of the CFT is $c = 6N$.

\begin{figure}[t!]
\centering
\begin{tabular}{ccc}
\includegraphics[width=.34\textwidth]{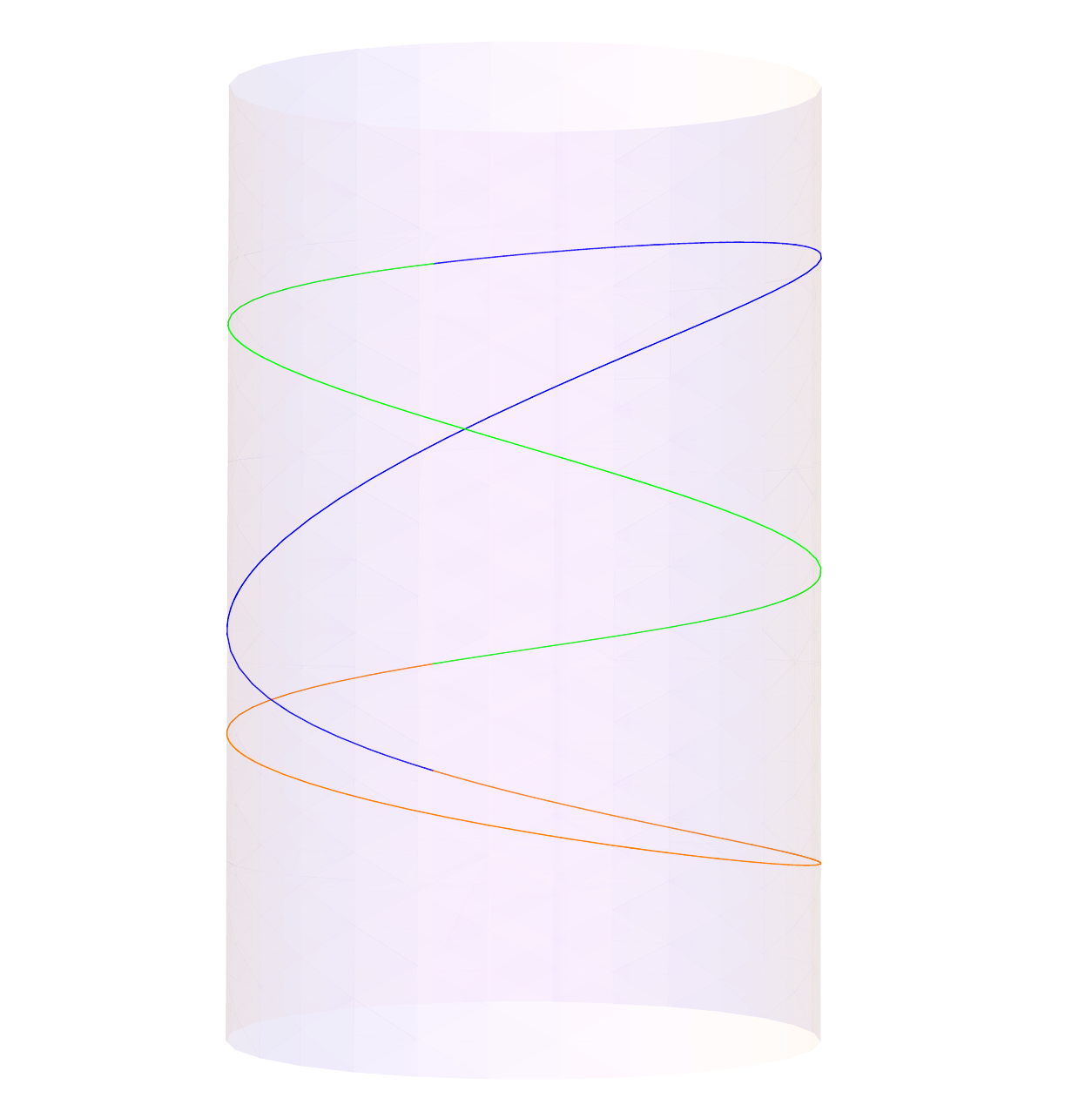} & \qquad\qquad\qquad &
\includegraphics[width=.25\textwidth]{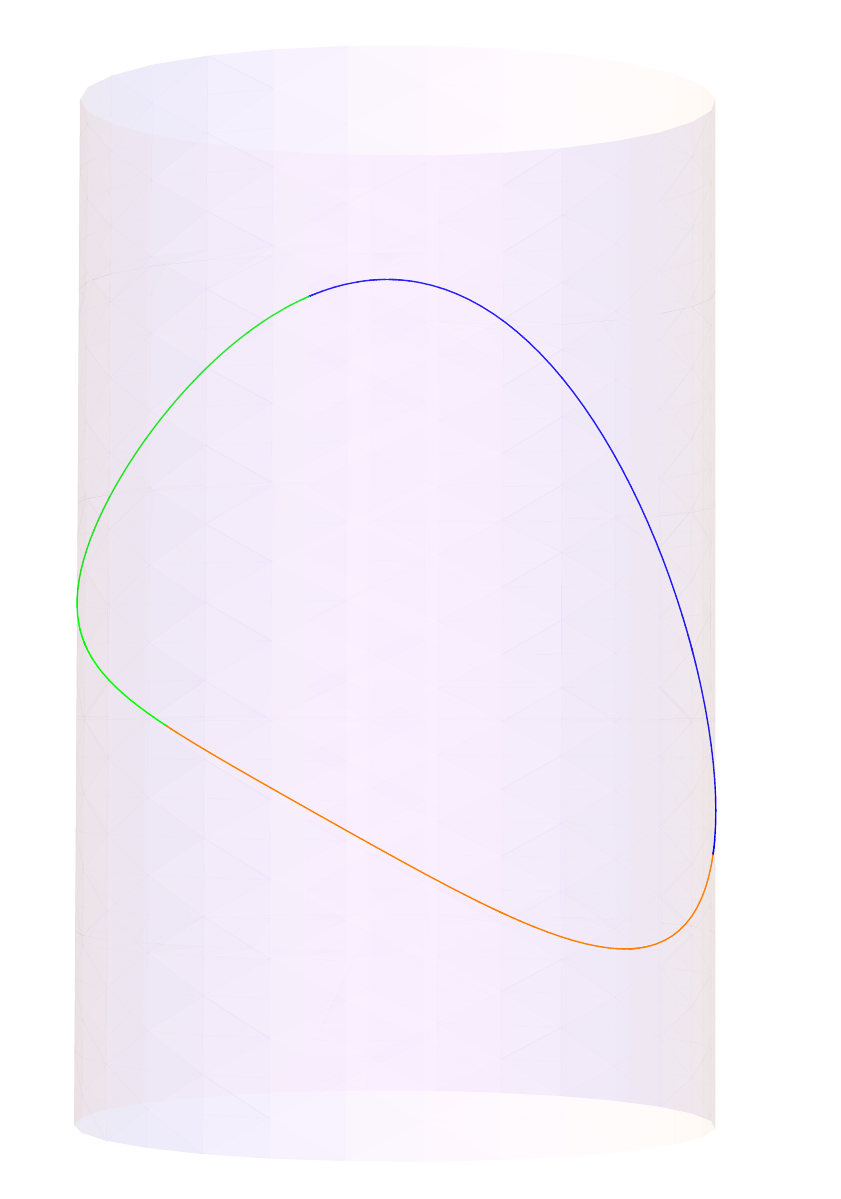}
\end{tabular}
\caption{A strand of three target space fields $X^{1,2,3}$, which define a single field $\tilde{X}^1$ of the long string.}
\label{gluing}
\end{figure}

We are interested in the geometry ${\rm AdS}_3/\mathbb{Z}_n$. Its conformal boundary is also (\ref{boundarygeom}), so this geometry should be dual to a state in the D1-D5 CFT. At the orbifold point this dual has been identified \cite{luninmathur, martinec, martinec2, vijaymasaki} as the state
\begin{equation}
(\sigma_{n})^{N/n} | 0 \rangle\,,
\label{state}
\end{equation}
which requires that $n$ be a divisor of $N$. Here $\sigma_n$ is a twist field. In this twisted sector, fields in the CFT are single valued on the $n$-fold cover of a spatial slice of the theory. More explicitly, a field configuration in the twisted sector is given by the profiles of $N$ $T^4$-valued target space fields around the worldvolume circle, which we call $X^1, \ldots, X^N$. The twisted boundary conditions set up by the twist fields in (\ref{state}) require that after a rotation by $2\pi$ around the $S^1$, the fields transform into one another as $N/n$ strands, each containing $n$ fields:
\begin{equation}
\begin{array}{rclcccrcl}
X^1 & \to & X^2 & & \ldots & & X^n & \to & X^1 \\
X^{n+1} & \to & X^{n+2} & & \ldots & & X^{2n} & \to & X^{n+1} \\
& & & & \ldots & & & & \\
X^{N-n+1} & \to & X^{N-n+2} & & \ldots & & X^N & \to & X^{N-n+1}
\end{array}
\label{twistedfields}
\end{equation}
This means that we can equivalently represent the theory with $N/n$ single-valued fields on a circle, whose circumference is $n$ times longer than the circle supporting the orbifold CFT.   We shall call such fields $\tilde{X}^j$, where $1 \leq j \leq N/n$. The values of $\tilde{X}^j$ are obtained by gluing together the values of the fields in~(\ref{twistedfields}), as illustrated in Fig.~\ref{gluing}. They define the worldvolume theory of the {\it long string} \cite{juanlenny}.\footnote{According to the standard orbifold prescription, we should still mod out the Hilbert space of the $N/n$ long strings by the the action of $\mathbb{Z}_n$ on each of the short strings, and also by the action of the permutation group $S_{N/n}$ which exchanges the long strings among themselves.  This is consistent with the covering CFT still being an orbifold theory, but now one based on $S_{N/n}$.}   So long as we consider untwisted probes and excitations, the dynamics is restricted to the superselection sector of this long string.  The construction of the long string on the $n$-fold cover of the short string means that we can think of the long string as living on the boundary of the covering space of the defect, which, as we discussed above, is simply empty AdS$_3$ space. Imposing the Brown-Henneaux relation \cite{BrownHenneaux}, we think of this AdS$_3$ cover as having a rescaled Planck constant $\tilde{G}$ \cite{JanJoanShahin}.

Because the long string is $n$ times longer than the short string, momentum on its worldvolume is quantized in units of $(nL)^{-1}$ instead of $L^{-1}$.    The same ratio applies to the spacings of the energy levels. The reduction in energy gap is called fractionation.   This reproduces our observation above based on the Virasoro algebras of the CFT$_c$ and CFT$_{\tilde{c}}$.  The factor $n$ also relates the central charges of the two theories: encapsulating the fields $X^1, \ldots, X^n$ in a single field $\tilde{X}^1$ trades $n$ degrees of freedom for a single degree of freedom, which leads to $\tilde{c} = c/n$.  This gives an explicit picture of the relation between $c$ and $\tilde{c}$ that we derived above from symmetry considerations and verified using entanglement computations. 
 The long string with central charge $\tilde{c} = c/n$ retains the information that it came from the state~(\ref{state}) of CFT$_c$ through the restriction on its set of gauge invariant observables and the action of $\mathbb{Z}_n$
on its Hilbert space.

\subsection{Entwinement and entanglement shadows}
\label{microent}

In Sec.~\ref{sec:entshadow} we saw that the conical defects have an entanglement shadow -- a central region which is not probed  by CFT entanglement.  This shadow exists because the minimal geodesics in the RT formula for holographic entanglement only penetrate to a certain maximum  depth in the spacetime.  The $n-1$ long geodesics (that go the long way around the defects or wind around it) do penetrate the entanglement shadow, but they do not contribute to the entanglement entropy, at least at the leading order, according to the RT formula.   Nevertheless, the long geodesics are certainly related to physical quantities in the CFT.  As discussed in Sec.~\ref{orbifoldcft}, they make sub-leading contributions to boundary correlation functions, and are in fact necessary for  conformal invariance.    Thus, we may wonder whether the long geodesics should also make a subleading contribution to the entanglement entropy (thereby modifying the entanglement shadow), perhaps via simple additive pieces resembling their method-of-images contributions to semiclassical correlation functions.  Such a picture is too simplistic if,  as discussed above, the conical defect can be regarded as a pure excited state of the D1-D5 string.  In this case, as the CFT interval tends to the size of the entire boundary, the entanglement entropy must tend to zero, which it will not if we include contributions from  the long geodesics (e.g. from the geodesics that start at a point, wind around the defect, and return to the same point).

The covering space picture in Fig.~\ref{rtincover} further illuminates the problem.   As discussed in Sec.~\ref{orbifoldcft}, an interval $\mathcal{R}$ in CFT$_c$ lifts to the union of $n$ evenly spaced intervals $ \tilde{R} = \cup_{i=1}^n  \tilde{\mathcal{R}}_i$ of size $2\tilde{\alpha} = 2\alpha / n$ in CFT$_{\tilde{c}}$.   So long as $2\tilde{\alpha} < \pi/n$, the minimal geodesics in the covering space  that are homologous to $\tilde{R}$ are arcs subtending the $\tilde{R}_i$.   But when $\pi/n < 2\tilde{\alpha} < 2 \pi/n$, the minimal geodesics transition to subtend the complementary intervals of angular size $(2\pi - 2\alpha)/n$ between the $\tilde{R}_i$,  precisely when the intervals $\tilde{R}_i$ begin to share mutual information.    If not for this mutual information transition in the covering space theory, the geodesics subtending the $\tilde{R}_i$ would have descended to  long geodesics in the conical defect when $\pi/n < 2\tilde{\alpha} <  2\pi/n$.   Once  the angular size of the $\tilde{R}_i$ is $2\pi/n$, their union covers the entire boundary and there is no entanglement entropy to be considered since the state is pure.   By contrast, {\it single} boundary intervals $\tilde{R}_i$ in the covering space of size  $\pi/n < 2\tilde{\alpha} < 2\pi  - \pi/n$ are spanned by minimal AdS$_3$ geodesics that will descend to long geodesics on the conical defect.    

We see that from the covering space perspective long geodesics are eliminated in the  holographic spatial entropy formulae by a mutual information transition that arises, because spatial entanglement entropy in the conical defect is computed from the entanglement of a $\mathbb{Z}_n$-invariant {\it union} of intervals in the covering space theory, i.e.
\begin{equation}
S(\mathcal{R}) 
= S\left( \cup_{i=1}^n \tilde{\mathcal{R}}_i \right) 
= S\left( \cup_{i=0}^{n-1} g^i \tilde{\mathcal{R}}_1 \right) 
\, 
\end{equation}
where $g$ is a $\mathbb{Z}_n$ generator (see Fig.~\ref{rtincover}).  Physical observables that descend to the conical defect certainly must be $\mathbb{Z}_n$-invariant.    Are there $\mathbb{Z}_n$-invariant quantities related to entanglement in CFT$_{\tilde{c}}$ that can be computed without first taking the $\mathbb{Z}_n$-invariant union of intervals in that theory?  One possibility is to compute the entanglement entropy of a single interval and then sum the result over $\mathbb{Z}_n$ translations:
\begin{equation}
E(\mathcal{R}) = \sum_{i=1}^n S\left( \tilde{\mathcal{R}}_i \right) 
=  \sum_{i=0}^{n-1} g^i S\left(\tilde{\mathcal{R}}_1 \right) 
\, .
\label{entwinementdef}
\end{equation}
This quantity, which we call {\it entwinement}, is $\mathbb{Z}_n$ invariant and thus  descends to the conical defect. If we make the intervals $\tilde{R}_i$ bigger than $\pi/n$, each term in (\ref{entwinementdef}) is separately computed in the covering AdS$_3$ by a minimal curve, which descends in the conical defect to a long geodesic that penetrates the entanglement shadow (see Fig.~\ref{geodesicspic}).  Explicitly, for a region $\mathcal{R}$ of total size $2\alpha$ we have:
\begin{equation}
E(\mathcal{R}) = n \cdot \frac{2L}{4\tilde{G}} \log \left( \frac{2L}{\mu} \sin\tilde\alpha \right) = \frac{c}{3} \log \left( \frac{2L}{\mu} \sin(\alpha/n) \right) = \frac{1}{4G}\, l(\alpha).
\label{entwexpl}
\end{equation}
We have used $3L/2\tilde{G} = \tilde{c} = c/n$ in a manner analogous to eq.~(\ref{bothcals}).

Let us summarize the steps we have taken to define entwinement.
We ``ungauged'' the discrete $\mathbb{Z}_n$ symmetry of the  conical defect theory, computed conventional spatial entanglement in the parent theory, and then symmetrized the computation to get a $\mathbb{Z}_n$-invariant quantity we called entwinement.    How can we interpret this quantity directly within the conical defect theory?  Recall that we argued that CFT$_c$, which is dual to the conical defect, has a set of ``internal'' degrees of freedom with fractionated energies and momenta.   We propose that entwinement captures the entanglement of subsets of these degrees of freedom in given spatial regions with the rest of the theory.   This interpretation is easiest to visualize in the long-string picture of the dual, which appears close to the orbifold point.  As described in Sec.~\ref{orbifoldcft}, in this picture there is an effective string with central charge $\tilde{c} = c/n$, which wraps $n$ times around the spacetime boundary and 
 hence has a total of $c$ degrees of freedom at each point.        Entwinement computes the closest analog to entanglement that applies to a partition of the windings into subsets.   If we compute the entwinement for intervals with $2\alpha < 2\pi$, we are calculating the entwinement of a part of one winding, summed over windings (i.e. summed over $\mathbb{Z}_n$ translations).   In the range $2\pi < 2\alpha < 4\pi$ we are computing the entwinement of between one and two windings of the effective string, summed over $\mathbb{Z}_n$ translations.    Our procedure of removing the $\mathbb{Z}_n$ identifications and  symmetrizing afterwards recalls methodologies that have been used before to study conventional entanglement entropy in gauge theories (see \cite{donnelly, entmaxwell, casinihuertaongaugeent, yuji} and references therein).

To emphasize the role of ungauging from a slightly different angle, consider a Hilbert space ${\cal H}={\cal H}_1 \otimes {\cal H}_2$ on which a group $G$ acts. One possibility is that $G$ maps ${\cal H}_1$ and ${\cal H}_2$ into themselves. This is the  case if, for example, ${\cal H}_1$ is the Hilbert space associated to the union of short intervals in the long string (whose associated geodesics descend to minimal geodesics in the conical defect) and ${\cal H}_2$ is the Hilbert space associated to the complement.  Even though $G$ preserves ${\cal H}_{1,2}$, this tensor factor decomposition does not descend to a tensor factor decomposition of ${\cal H}^G$. We can certainly decompose ${\cal H}_1$ and ${\cal H}_2$ into irreps $R_i$ of $G$, so that the decomposition reads
 \be
{\cal H} = \oplus_{i,j} {\cal H}_1^{R_i} \otimes {\cal H}_2^{R_j}.
\ee
Now ${\cal H}^G$ will only contain $G$-singlets, and therefore only contains contributions from the sum when the representations are conjugate, $R_i=\bar{R_j}$, and even then one still has to project in general on $G$-invariant states. One is always left with a complicated sum of tensor factors. Thus, even for a short interval, the appropriate notion of entropy cannot be obtained as the entanglement entropy associated to a tensor factor in the invariant Hilbert space ${\cal H}^G$  and one always needs to ungauge.

For sufficiently long intervals, associated to long geodesics in the conical defect, there is generally not a decomposition ${\cal H}={\cal H}_1 \otimes {\cal H}_2$ of the long string Hilbert space where $G$ preserves ${\cal H}_{1,2}$. What we have effectively done is to pass to an even bigger Hilbert space
 \be
 {\cal H}_{\rm extended} = \oplus_{g\in G} g({\cal H}_1)\otimes g({\cal H}_2)
 \ee
whose decomposition is now compatible with an action of $G$ (it permutes the summands). It would clearly be interesting to explore the connection between such decompositions and the discussion in \cite{donnelly, entmaxwell, casinihuertaongaugeent, yuji} (and references therein).

\section{Reconstruction of geometry from entwinement}

\subsection{Geometry from entanglement}
\label{reconstruction}

\begin{figure}[p]
\centering
\vspace*{-1.6cm}
\includegraphics[width=.35\textwidth]{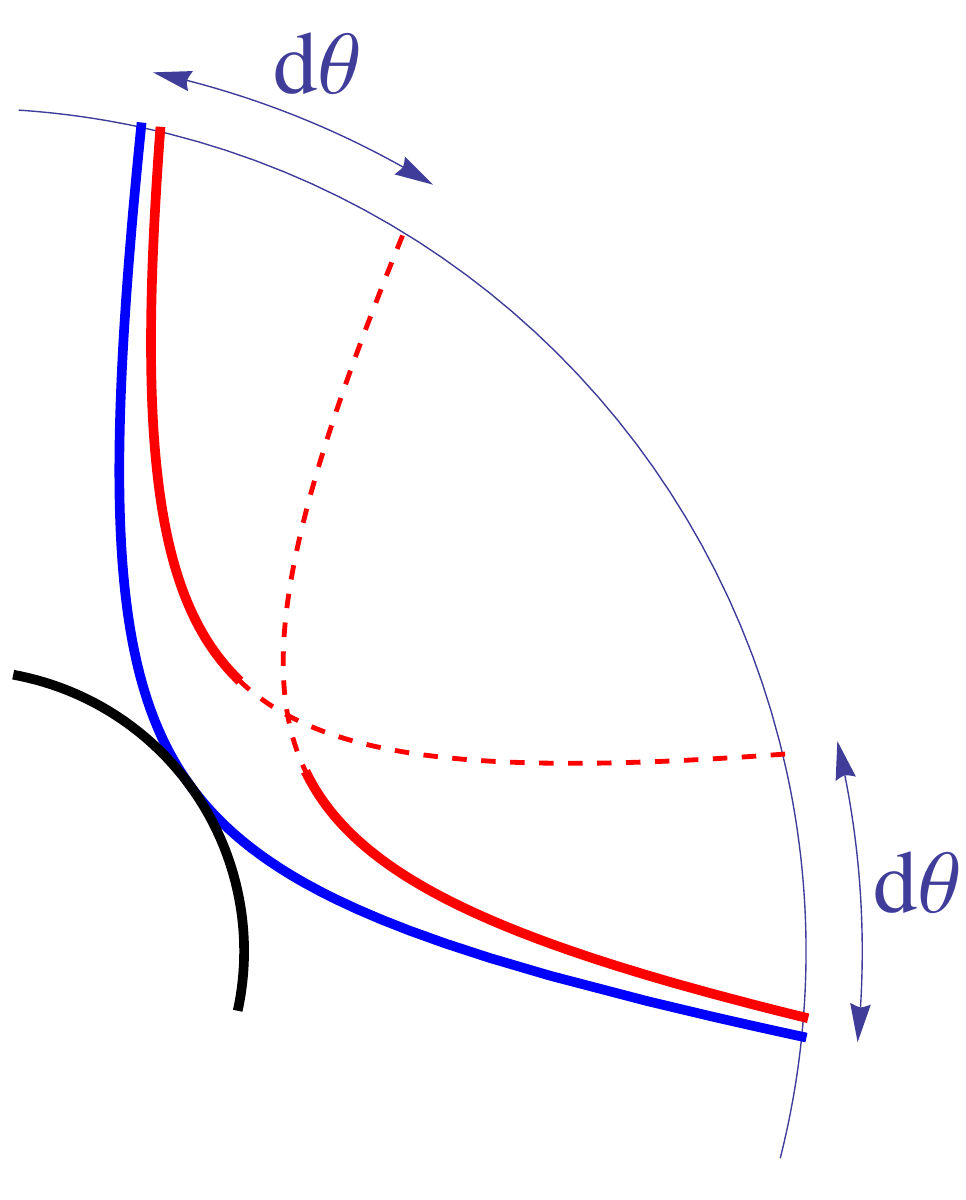}
\captionsetup{singlelinecheck=off}
\caption[]{
The integrand of eq.~(\ref{adscurve0}) as a finite difference:
\allowbreak    
\begin{equation*}
\frac{1}{2}\frac{d\,l(\alpha)}{d\alpha}\,d\theta \approx l(\alpha) - l(\alpha-d\theta/2) = {\color{blue} l(\alpha)} 
- \frac{1}{2}\, {\color{red} l(\alpha-d\theta/2)} - \frac{1}{2}\, {\color{red} l(\alpha-d\theta/2)}
\approx \frac{R_0 \,d\theta}{4G}
\end{equation*}
The color-coded summands correspond to the continuously drawn pieces of the geodesics in the figure. The difference between their lengths aligns with the length element along the central circle $R = R_0$.
\vspace*{0.7cm}
}
\label{lengthelement}
\includegraphics[width=.51\textwidth]{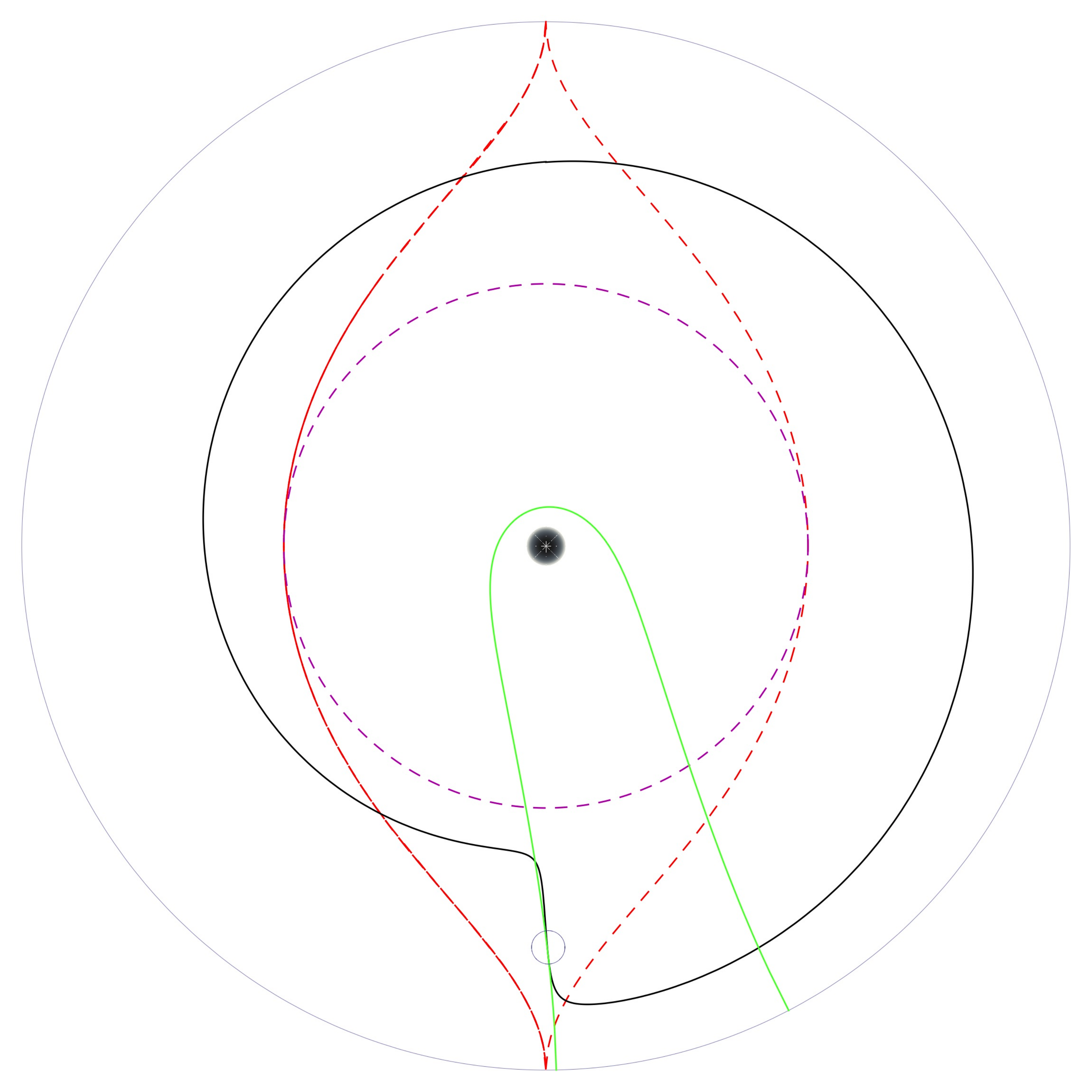}
\caption{A closed curve (black) in the conical defect spacetime, which probes entwinement. The long geodesic tangent to it is shown in green and the tangency point is marked with a circle. Any curve that is steeper (approaches the boundary faster) than the marginal geodesics ($\alpha = \pi/2$, shown in dashed red) is tangent to a long geodesic and therefore probes entwinement. In particular, every curve that is locally parallel to the radial direction probes entwinement. In the entanglement shadow ($r < r_{\rm min}$, marked in dashed purple) all curves probe entwinement, regardless of the slope.}
\label{wheremicro}
\end{figure}

In \cite{lastpaper}, we showed how to compute the circumference of an arbitrary, piecewise differentiable closed curve on a spatial slice of 2+1-dimensional anti-de Sitter space from boundary data. For definiteness, we work in global coordinates
\begin{equation}
ds^2 = - \left( 1 + \frac{R^2}{L^2} \right) dT^2 + \left( 1 + \frac{R^2}{L^2} \right)^{-1} dR^2 + R^2 d\tilde\theta^2
\end{equation}
and represent the curve with the equation $R = R(\tilde\theta)$. To every point on the curve one associates a boundary interval $\mathcal{I}(\tilde\theta)$ of length $2\alpha(\tilde\theta)$ centered at $\theta(\tilde\theta)$. The interval can be determined in one of two ways:
\\
\parbox[t]{0.25 in}{ }
\ \hspace{0.15 in} 
\parbox[t]{6.25in}{
 {\bf (1)} The outgoing null ray orthogonal to the curve reaches the boundary after a global time $L \alpha(\tilde\theta)$ at the spatial location $\theta(\tilde\theta)$. 
\\
 {\bf (2)} The spatial geodesic tangent to the curve at $\tilde\theta$ subtends the interval $\mathcal{I}(\tilde\theta)$, that is it connects boundary points $\theta(\tilde\theta) \pm \alpha(\tilde\theta)$.
}\\
In a spacetime that is locally AdS$_3$, the two definitions of $\alpha(\tilde\theta)$ and $\theta(\tilde\theta)$ are equivalent. Then the length of the closed curve is given by the formula:
\begin{equation}
\frac{\rm length}{4G} = 
\frac{1}{4G}\cdot \frac{1}{2}\int_0^{2\pi} d\theta \,\,\frac{d\, l(\alpha)}{d\alpha} 
\Bigg|_{\alpha=\alpha(\theta(\tilde\theta))} = \frac{1}{2} \int_0^{2\pi} d\theta \,\,\frac{d\, S_{\rm ent}(\mathcal{I}(\tilde\theta))}{d\alpha} \Bigg|_{\alpha=\alpha(\theta(\tilde\theta))}
\label{adscurve0}
\end{equation}
The second equality uses the Ryu-Takayanagi relation~(\ref{eqrt}). In the special case of a central circle, $R=R_0$, one can interpret the integrand $dl(\alpha)/d\alpha|_{\tilde\theta}$ as corresponding to an infinitesimal length element along the curve at $\tilde\theta$. This reasoning, which is illustrated in Fig.~\ref{lengthelement}, explains formula~(\ref{adscurve0}) from the bulk point of view. For a general bulk curve, a similar explanation holds, though the technical details are more involved. For more information, consult~\cite{lastpaper}.

\subsection{Reconstructing the conical defect spacetime}
\label{reconst}

The conical defect spacetime is locally AdS$_3$. Therefore, the middle formula in eq.~(\ref{adscurve0}), which is an identity in the bulk, extends to the conical defect automatically. The version on the right hand side, however, is a boundary statement, which applies only so long as
\begin{equation}
S_{\rm ent}(\mathcal{I}(\tilde\theta)) = \frac{1}{4G}\,l(\alpha(\tilde\theta)).
\end{equation}
As we saw in eqs.~(\ref{shortcal}-\ref{longercal}), this requires that $\alpha \leq \pi/2$. In the opposite case, that is when construction {\bf (2)}  in Sec.~\ref{reconstruction} above returns a long geodesic, the set of spatial boundary entanglement entropies is insufficient to define the given bulk curve, let alone to calculate its length. As we saw in eq.~(\ref{defmin}), this occurs whenever the curve approaches the conical defect AdS$_3/\mathbb{Z}_n$ closer than a coordinate distance $r_{\rm min} = L \cot (\pi/2n) / n$. In this regime, the entwinement computed in eq.~(\ref{entwexpl}) is a necessary ingredient.

In fact, long geodesics and entwinement are also necessary to recover the metric outside this central zone. In particular, any closed curve with a sufficiently `radial' local direction gives rise to $\alpha(\tilde\theta) > \pi/2$, as is illustrated in Fig.~\ref{wheremicro}. The critical direction as a function of radial scale is set by the marginal geodesic $\alpha = \pi/2$, which separates long from short geodesics:
\begin{equation}
\tan^2(\theta/n) = \frac{n^2 r^2 \tan^2(\pi/2n) - L^2}{n^2 r^2 + L^2} 
\label{criticalgeod}
\end{equation}
A curve, which at any radial scale approaches the boundary more rapidly than geodesic~(\ref{criticalgeod}), cannot be defined or measured in the boundary using spatial entanglement entropies alone.  (See the related discussion in \cite{myersetal}.)  Importantly, without this information one cannot determine the radial component of the bulk metric anywhere. The entanglement shadow in the region $r < r_{\rm min}$ is special in that there even the angular component of the metric is inaccessible from boundary spatial entanglement entropies.

As an example, consider the central circle $r = r_0$ in metric~(\ref{defmetric}). Using method {\bf (2)}  from Sec.~\ref{reconstruction} and the geodesic~(\ref{defgeodesics}), we find $\alpha(\tilde\theta) = n \tan^{-1} (L/nr_0)$. Plugging its length (\ref{deflength}) into the middle expression in eq.~(\ref{adscurve0}), we obtain:
\begin{equation}
\frac{1}{4G}\cdot \frac{1}{2}\int_0^{2\pi} d\theta \,\,\frac{d\, l(\alpha)}{d\alpha} 
\Bigg|_{\alpha=n \tan^{-1} (L/nr_0)} = 
\frac{1}{4G}\cdot \frac{1}{2}\int_0^{2\pi} d\theta \,\frac{2L}{n} \cot\frac{\alpha}{n}\Bigg|_{\alpha=n \cot^{-1} (nr_0/L)}
= \frac{2\pi r_0}{4G}
\label{smallcirc}
\end{equation}
When $r_0 < r_{\rm min}(n)$ (see eq.~\ref{defmin}), $\alpha > \pi/2$ and the holographic interpretation of $l(\alpha)$ is given in eq.~(\ref{entwexpl}) in terms of entwinement. More generally, entwinement is necessary to define and measure any curve with a sufficiently radial direction anywhere, see Fig.~\ref{wheremicro}.

\section{Discussion}

We have argued that entwinement -- an analog of entanglement for gauged Hilbert spaces  which do not contain all the degrees of freedom of spatial regions in a field theory -- is an essential ingredient for holographically reconstructing the conical defect geometries in three dimensions from field theory data.  There is a central zone near the conical defect, which is not probed by conventional spatial entanglement in the field theory and entwinement becomes necessary.   Furthermore, even far away from the defect, entwinement is implicated in the the emergence of the radial direction of space.  

One of our lines of argument involved the orbifold description of the weakly coupled D1-D5 string, which made the discrete $\mathbb{Z}_n$ gauge symmetry easy to visualize and manipulate.    The field theories that we encounter in AdS/CFT are usually low energy limits of gauge theories, and we therefore expect that some notion of gauge invariance is still present away from the weak coupling limit. This is illustrated directly by the CFT that is dual to the covering space description of the conical defect, and which is the strongly coupled version of the long string sector that is evident at weak coupling. Therefore, we expect the existence of a field-theoretic definition of entwinement away from weak coupling as well.

The long geodesics associated to entwinement map to spatial intervals that cover the boundary of the conical defect more than once, suggesting that the associated observers have access to all the information available in the field theory.   However, following method {\bf (1)} of Sec.~\ref{reconstruction}, one can show that the associated time intervals on the boundary are finite.   It was proposed in \cite{lastpaper} that ignorance of the quantum state associated to such finite time measurements (called ``residual entropy'' in \cite{lastpaper}) is related to the areas of closed curves in the bulk spacetime and their associated entropies \cite{holeinspacetime}.  Indeed, in the present case, the observers associated to  time intervals that are too short will not have the energy resolution to probe excitations with the fractionated gaps of the conical defect.   More precisely, it is clear from  the bulk point of view (employing the covering CFT), as well as from the long string point of view, that by causality such observers cannot fully access all $\mathbb{Z}_n$ invariant correlation functions. Entwinement is a quantity which might be associated to this ignorance and it would be very interesting to make the connection more precise.  One can perhaps get some clues from recent work by Hubeny \cite{Hubeny:2014qwa}, which investigates potential covariant definitions of the residual entropy associated to finite time observers, and possible relations (or the lack thereof) to minimal holographic surfaces.

\subsection{Beyond conical defects}

We believe that these lessons are not limited to conical defect geometries, but apply in greater generality.  Consider, for example, the massless BTZ geometry \cite{BTZreview} whose metric is the $n\to\infty$ limit of the conical defect metric~(\ref{defmetric}). Our results apply for any $n$, which divides $N = N_1 N_5$ (see Sec.~\ref{orbifoldcft}), but we anyway presume that $N$ approaches infinity whenever we discuss geometric quantities in the bulk. Thus, the results of the present paper extend directly to the massless BTZ geometry, except we must take $n = \infty$. In particular, the massless BTZ spacetime contains spatial geodesics that wrap around the black hole infinitely many times, as opposed to the maximal number of $n/2$ for AdS$_3/\mathbb{Z}_n$ (see eq.~\ref{deflength}). The entanglement shadow, which short geodesics do not reach, extends out to:
\begin{equation}
r_{\rm crit}(\infty) = \lim_{n \to \infty} r_{\rm crit}(n) = \frac{2L}{\pi}.
\end{equation}
The same critical radial scale can be found directly from the form of spatial geodesics
\begin{equation}
r^2 = r_+^2\, \frac{\cosh^2(r_+ \alpha/L)}{\sinh^2(r_+ \alpha/L) - \sinh^2(r_+ \theta/L)}
\label{BTZgeod}
\end{equation}
in the massive, stationary BTZ metric
\begin{equation}
ds^2 = -\left(\frac{r^2 - r_+^2}{L^2}\right) \,dt^2 
+ \left(\frac{L^2}{r^2 - r_+^2}\right) dr^2 + r^2 d\theta^2.
\label{btzmetric}
\end{equation}
by taking the limit $r_+ \to 0$.\footnote{In the massive BTZ spacetime, the range of spatial geodesics with $\alpha > \pi/2$ up to the so-called entanglement plateau \cite{plateaux} are also short in the sense that they compute entanglement entropies of spatial regions on the boundary. This is due to the homology constraint in the Ryu-Takayanagi proposal. But in the special case of the massless BTZ, the maximal opening angle of a short geodesic is $\pi/2$, as is evident from the Araki-Lieb inequality \cite{al}:
\begin{equation}
|S(\mathcal{R}) - S(\mathcal{R}^c)| \leq S(\mathcal{R}\cup\mathcal{R}^c) = S_{\rm BTZ}= 0.
\end{equation}}

Beyond the massless BTZ geometry, a qualitatively new ingredient appears. When we obtain the conical defect geometry as an orbifold of AdS$_3$, we mod out by a finite subgroup of rotations. Formally speaking, we orbifold by an elliptic element of the conformal group. The massive BTZ geometries are obtained from orbifolding by a hyperbolic element,  in other words, a boost.\footnote{The massless BTZ is the critical case of orbifolding by a parabolic element. As we saw in the previous paragraph, it can be recovered either as a limit of elliptic or hyperbolic orbifolds.} The argument presented in this paper identified the boundary of the AdS cover with the worldvolume theory of the long string. In the BTZ case, this identification is more complex.  First, the orbit of a boost is noncompact, so the covering space contains infinitely many copies of the BTZ spacetime. Second, the boost acts non-trivially on the time direction, which means that a $t=const.$ slice of the BTZ boundary lifts does not lift to a $T=const.$ slice of the boundary of the covering space. Tackling these complications is likely to teach us more about the emergence of horizons from entanglement in the dual boundary theory.\footnote{Note, however, that several authors \cite{BTZproblems1, BTZproblems2, BTZproblems3,BTZproblems4,BTZproblems5} have suggested subtleties that might challenge the interpretation of  the BTZ black hole as an orbifold at the quantum level.}

For the weakly coupled orbifold theory, we can extend our proposal to more general states and in particular thermal
states. The full Hilbert space then involves a sum over tensor products of strings of various lengths, and by summing
the entanglement of long intervals over those the twisted sectors that can accomodate long intervals with
appropriate weights, we obtain a natural generalization. It would be interesting to do this computation for a thermal
state and to compare to the lengths of long geodesics (i.e. non-minimal geodesics that wind around the horizon)  in the massive BTZ background.

\subsection{Spatial entanglement, entwinement and mutual information}
As formulated above for the case of the conical defect, the difference between spatial entanglement and entwinement lay in the order in which we carried out the $\mathbb{Z}_n$ symmetrization:
\begin{equation}
S\left( \cup_{i=1}^n \tilde{\mathcal{R}}_i \right) \neq \sum_{i=1}^n S\left(\tilde{\mathcal{R}}_i \right) \, .
\label{difference}
\end{equation}
This difference can be traced to the nonvanishing mutual information between the image regions $\tilde{\mathcal{R}}_i$. Specializing to the case $n=2$, the mutual information is precisely the difference highlighted in (\ref{difference}):
\begin{equation}
I(\tilde{\mathcal{R}}_1 : \tilde{\mathcal{R}}_2) = S(\tilde{\mathcal{R}}_1) + S(\tilde{\mathcal{R}}_2) - S(\tilde{\mathcal{R}}_1 \cup \tilde{\mathcal{R}}_2)
\label{mutualinfo}
\end{equation}
At least in the special case of AdS$_3/\mathbb{Z}_2$, the mutual information between images can therefore be thought of as an order parameter for the relevance of entwinement in a holographic reconstruction of spacetime.

\subsection{Toward higher dimensions}

In the conical defect geometry entwinement computes areas of extremal but nonminimal curves. In higher dimensions, extremal but nonminimal surfaces likewise play a role in the emergence of holographic spacetimes. One example was given in \cite{xijamieandi}, which generalized formula (\ref{adscurve0}) to compute areas of codimension-2 surfaces in higher-dimensional spaces. Like eq.~(\ref{smallcirc}), that computation generally involves extremal but nonminimal surfaces. It would be fascinating to lift the definition of entwinement, which in its present form pertains to 2d CFTs, to give a boundary interpretation of areas of extremal but nonminimal surfaces in higher dimensions.

We saw in Sec.~\ref{reconst} that entwinement is relevant in boundary computations of bulk lengths under two distinct circumstances: when the bulk curve becomes nearly radial and when it probes the region near the conical singularity. The computation in \cite{xijamieandi} is roughly analogous to the former circumstance, but does the latter have an analogue in higher dimensions? More specifically, when do higher-dimensional spacetimes possess entanglement shadows -- regions outside the reach of minimal surfaces?\footnote{We thank Mukund Rangamani and Mark Van Raamsdonk for clarifying this issue and for pointing out Refs.~\cite{fernando, mukund}.} Ref.~\cite{plateaux} showed that AdS-Schwarzschild black holes are similar in this respect to BTZ black holes: they have entanglement shadows that surround the horizon up to a thickness of order $L_{AdS}$. 
On the other hand, it is known that horizonless geometries with matter can have no entanglement shadow \cite{fernando, mukund}.

\section*{Acknowledgments}
We are grateful to Michal Heller for his collaboration in the early part of this project and for many discussions.  We thank Dionysios Anninos, Xi Dong, Liam Fitzpatrick, Patrik Hayden, Matthew Headrick, Jared Kaplan, Nima Lashkari, Javier Mag{\'a}n, Rob Myers, Mukund Rangamani, Joan Sim{\'o}n, James Sully, Mark Van Raamsdonk, Erik Verlinde, and Herman Verlinde for helpful discussions. VB was supported by DOE grant DE-FG02-05ER-41367 and by the Fondation Pierre-Gilles de Gennes. The work of BC has been supported in part by the Stanford Institute for Theoretical Physics. 
This work is part of the research programme of the Foundation for Fundamental Research on Matter (FOM), which is part of the Netherlands Organisation for Scientific Research (NWO). BC dedicates this paper, submitted on Father's Day, to his Dad, Stanis{\l}aw Czech.

\end{document}